\documentclass[
reprint,     
superscriptaddress,
showkeys,
 amsmath,amssymb,
 aps,
prl,
floatfix,
]{revtex4-2}

\usepackage{natbib}
\usepackage{graphicx}
\usepackage{dcolumn}
\usepackage{bm}

\usepackage{tikz}
\usepackage{tikz-cd}
\usepackage{pgfplots}
\pgfplotsset{compat=1.14}

\usepackage{svg}

\usepackage[hidelinks]{hyperref}
\usepackage[mathlines]{lineno}

\graphicspath{{figs/}{figures/}{figures_new/}}
\usepackage{float} 
\begin{document}

\title{Broadband limits on stochastic length fluctuations from a pair of table-top interferometers}

\author{Abhinav Patra}
\email{patraa1@cardiff.ac.uk}
\affiliation{Gravity Exploration Institute, Cardiff University, Cardiff CF24 3AA, United Kingdom}

\author{Lorenzo Aiello}
\thanks{Current affiliation - Università di Roma Tor Vergata, I-00133 Roma, Italy and INFN, Sezione di Roma Tor Vergata, I-00133 Roma, Italy.}
\affiliation{Gravity Exploration Institute, Cardiff University, Cardiff CF24 3AA, United Kingdom}

\author{Aldo Ejlli}

\affiliation{Gravity Exploration Institute, Cardiff University, Cardiff CF24 3AA, United Kingdom}

\author{William L. Griffiths}

\affiliation{Gravity Exploration Institute, Cardiff University, Cardiff CF24 3AA, United Kingdom}

\author{Alasdair L. James}

\affiliation{Gravity Exploration Institute, Cardiff University, Cardiff CF24 3AA, United Kingdom}

\author{Nikitha Kuntimaddi}

\affiliation{Gravity Exploration Institute, Cardiff University, Cardiff CF24 3AA, United Kingdom}

\author{Ohkyung Kwon}
\affiliation{University of Chicago, 5640 S. Ells Ave., Chicago, IL 606035, USA}
\affiliation{Gravity Exploration Institute, Cardiff University, Cardiff CF24 3AA, United Kingdom}

\author{Eyal Schwartz}

\affiliation{Gravity Exploration Institute, Cardiff University, Cardiff CF24 3AA, United Kingdom}

\author{Henning Vahlbruch}
\affiliation{Max-Planck-Institute for Gravitational Physics and Leibniz University Hannover, Callinstr. 38, 30167 Hannover, Germany}

\author{Sander M. Vermeulen}

\affiliation{Gravity Exploration Institute, Cardiff University, Cardiff CF24 3AA, United Kingdom}

\author{Keiko Kokeyama}
\affiliation{Gravity Exploration Institute, Cardiff University, Cardiff CF24 3AA, United Kingdom}

\author{Katherine L. Dooley}
\affiliation{Gravity Exploration Institute, Cardiff University, Cardiff CF24 3AA, United Kingdom}

\author{Hartmut Grote}
\affiliation{Gravity Exploration Institute, Cardiff University, Cardiff CF24 3AA, United Kingdom}

\date{\today}

\begin{abstract}
The Quantum-Enhanced Space-Time (QUEST) experiment consists of a pair of co-located Power Recycled Michelson Interferometers, each designed to have a broadband, shot-noise limited displacement sensitivity of $2\times10^{-19}$ $\mathrm{m/\sqrt{Hz}}$ from 1 to 200\,MHz. Here we present the first results of QUEST, with a search up to 80\,MHz, that set new upper limits on correlated length fluctuations from 13 to 80\,MHz, constituting the first broadband constraints for a stochastic gravitational wave background at these frequencies. In a coincident observing run of $10^{4}$\,s the averaging of the cross-correlation spectra between the two interferometer signals resulted in a strain sensitivity of $3\times10^{-20}$  $\mathrm{1/\sqrt{Hz}}$, making QUEST the most sensitive table-top interferometric system to date.

\end{abstract}


\maketitle

The development of laser interferometry over the last decades has allowed for frequent detection of gravitational waves, revolutionizing astrophysics~\cite{PhysRevX.13.041039}. So far, all detected gravitational-wave signals are from non-stationary events, understood as signatures from the coalescence of compact astronomical objects such as black holes and neutron stars. However, gravitational waves are also expected from stationary sources such as rotating neutron stars (in the form of quasi-continuous waves) and from the stochastic superposition of unresolved astrophysical or cosmological sources, such as primordial black holes, cosmic string loops, and other relics possibly produced in the early universe \cite{10.1093/mnras/168.2.399,PhysRevLett.98.111101,PhysRevD.109.103538,Cruise_2012,Aggarwal_2021,Mingarelli_2019}. Other sources of stationary signals that can be interrogated with laser interferometry are, for example, signatures of scalar field dark matter \cite{Grote:2019uvn,Vermeulen:2021epa,Aiello:2021wlp,Gottel2024} and possible signatures of quantum gravity in the form of stochastic space-time fluctuations \cite{Hogan:2008a,Kwon:2014yea,Kwon_2025,VERLINDE2021,BanksZurek2021,Verlinde2022,banks2023fluctuations,Sharmila2023, PhysRevD.109.123505}.

A powerful technique to search for stationary stochastic signals that manifest as length fluctuations is to facilitate cross-correlation between two laser interferometers that share the scientific target signal but are commissioned to be independent in their technical noises ~\cite{H1H22015}. The Quantum-Enhanced Space-Time (QUEST) experiment is dedicated to develop correlated interferometry for fundamental physics and targets signals such as quantum gravity, dark matter, and high-frequency gravitational waves ~\cite{vermeulen_experiment_2021}. The Holometer experiment at Fermilab, pioneered this technique and demonstrated the feasibility of the method using 40-meter scale interferometers, while the QUEST and the planned GQuEST \cite{vermeulen2024photoncountinginterferometrydetect} experiments are developing the technology further and probing broader frequency bands using table-top scale instruments.

In this letter, we report on the first results of the QUEST experiment that demonstrate the feasibility of co-located table-top interferometry by providing new upper limits on stationary length fluctuations. These results constitute new constraints on stochastic gravitational waves in the frequency band from 13 to 80\,MHz, improving upon the limits established by previous interferometric experiments in this frequency band~\cite{Akutsu2008, Chou2017}.

Similar to long-baseline interferometers such as the Laser Interferometer Gravitational wave Observatories (LIGO), the interferometers in QUEST are operated as linear phase detectors, allowing for measurement of signals that cause relative phase accumulation between the light fields traversing their arms. The interferometers in QUEST utilize a homodyne readout scheme with a feedback loop to lock the arm length difference of each interferometer to an offset near the true dark fringe, resulting in a fixed but small fraction of probe field from inside the interferometer to leak out to the output port. The fields exiting are continuously measured using photo-detectors to extract time-varying information of the accumulated phase difference between the arms.

The Heisenberg uncertainty principle bounds the accuracy to which the phase of the probe field can be measured as a time-varying quantity. In the absence of a limiting classical noise, in the frequency band of interest, this manifests as a counting uncertainty of the probe field photons known as photon shot noise. The uncertainty has a Poisson distribution over the expected number of photons in the circulating field, resulting in a frequency-independent noise.

When using an un-squeezed vacuum input to the anti-symmetric port of an interferometer, the shot noise limited sensitivity towards changes to the differential-arm length degree of freedom is given by \cite{grote_text, caves81} - 

\begin{align}
    S^{\mathrm{sn}}(f) &= \sqrt{\frac{c\hbar \lambda}{4\pi P_{\mathrm{BS}}}} \\
    &\approx \bigg(5\times 10^{-19}\,{\mathrm{\frac{m}{\sqrt{Hz}}}}\bigg)\bigg(\frac{10\,\mathrm{kW}}{P_{\mathrm{BS}}}\bigg)^{1/2} \label{eq:sens}
\end{align}

\noindent where $\lambda$ is laser wavelength - 1064\,nm in QUEST - and $P_{\mathrm{BS}}$ is the laser power on the beamsplitter.

\begin{figure}[t]
    \includegraphics[width=8.6cm]{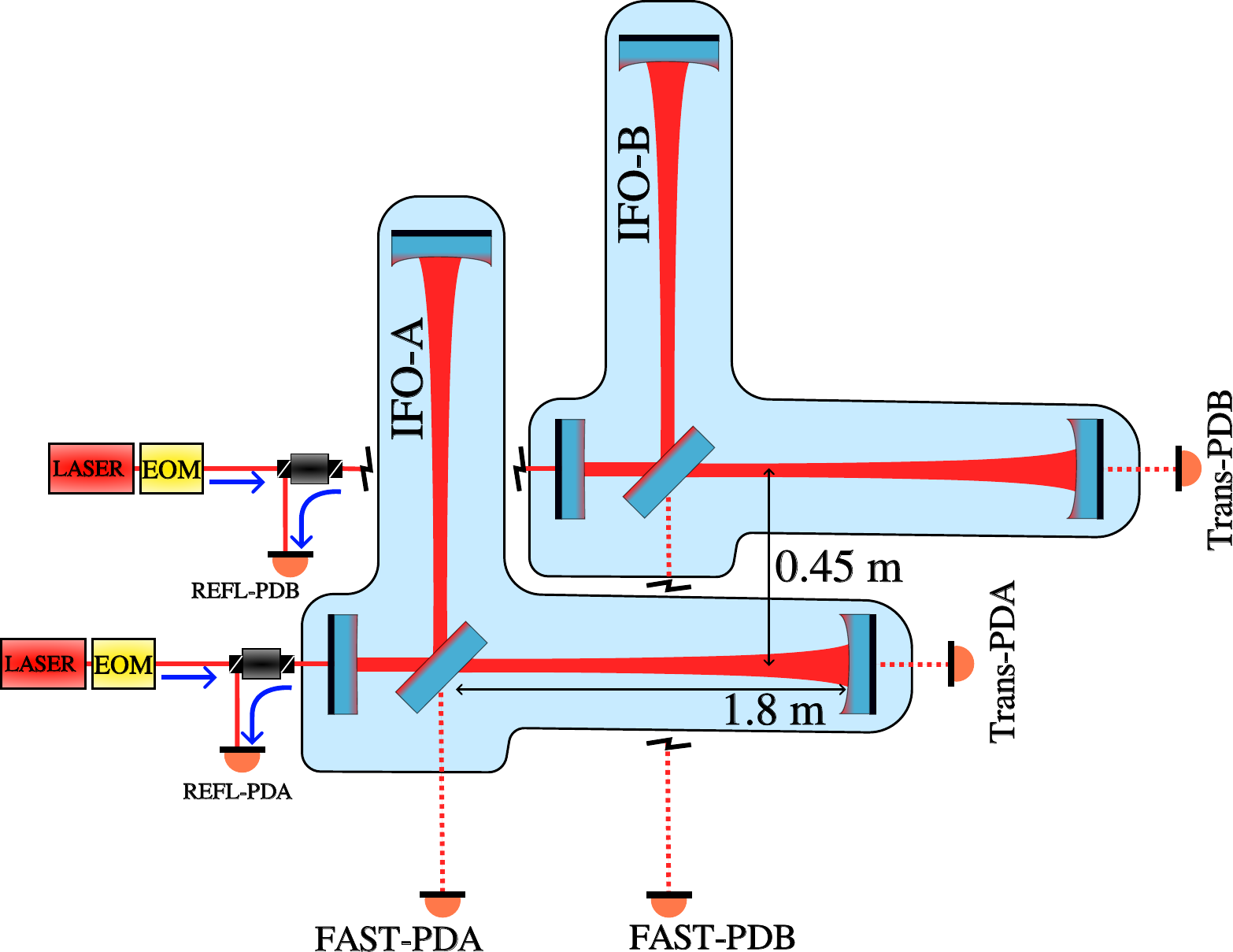}
    \caption{\label{fig:epsart} A  schematic of the interferometers in the QUEST experiment. The vacuum envelopes, core optics, laser sources and the main photo-detectors are shown. Electro-optic modulators are used to inject phase-modulation sidebands for the Pound-Drever-Hall sensing scheme.}
\end{figure}

\textit{Experimental setup -} QUEST consists of two co-located and co-aligned Power Recycled Michelson Interferometers with perpendicular arms. The common mode power-recycling resonantly enhances the effective power circulating in the arms, thus improving the shot noise-limited sensitivity of the interferometers. The interferometers in QUEST have an arm length of 1.83\,m and an inter-arm separation of 0.45\,m. The end mirrors of the interferometers are highly reflective (12.7\,ppm transmissivity) and the power recycling mirror has a reflectivity of 99.5\%. 
A simplified schematic of the experimental apparatus is shown in FIG.\,\ref{fig:epsart}. 

The two interferometers in QUEST are housed in two separate vacuum envelopes, that sit on a vibration isolated optical bench and each vacuum envelope is equipped with a set of mechanical turbo pumps to acquire vacuum. To maintain the vacuum levels and to avoid any vibrational noise being coupled to the interferometers via mechanical pumps, each vacuum envelope is also equipped with a set of non-mechanical ion pumps. When operating with just the ion pumps the vacuum levels settle at ultra-high vacuum pressures of about $7\times10^{-8}$\,mbar \cite{vermeulen_experiment_2021}.

The core optics of the interferometers - the power recycling mirror, the beamsplitter and the end test masses - are rigidly mounted inside the vacuum chambers using custom made optical mounts. Each core optic mount is equipped with pico-motors for coarse alignment and the end test masses are also equipped with piezo-electric actuators that give control over longitudinal position and angular pointing of the optics.

Two feedback loops are commissioned per interferometer to maintain the instruments at their operating point. A frequency stabilization loop, utilizing the Pound-Drever-Hall (PDH) technique, is used to lock the frequency of the laser to the length of the power recycling cavity and a differential arm length (DARM) stabilization loop is used to lock the arm length difference of the interferometer by utilizing the power exiting the output port (homodyne readout). The angular degrees of freedom are fine-tuned in lock to improve the contrast and power buildup of the interferometers.

\textit{Correlation analysis and averaging -} In QUEST, the target signals are assumed stationary and correlated between the two interferometers. Therefore, cross-correlation and averaging between the outputs of the interferometers can be performed to suppress uncorrelated technical and quantum noise. The photon-shot-noise dominated time series signal from each interferometer is Fourier transformed and a frequency domain cross-correlation is calculated. The cross-correlated spectra (CSD) are then averaged:

\begin{align}
    &S^{\mathrm{N}}_{\mathrm{AB}}(f) = \frac{1}{N}\sum_{n=1}^{\mathrm{N}} F_{\mathrm{A}}^{\dagger}(n, f)\cdot F_{\mathrm{B}}(n, f)\\
    &S_{\mathrm{AA}} = F_{\mathrm{A}}^{\dagger}\cdot F_{\mathrm{A}}; S_{\mathrm{BB}} = F_{\mathrm{B}}^{\dagger}\cdot F_{\mathrm{B}}
\end{align}

\noindent where $S_{\mathrm{AA}}$, $S_{\mathrm{BB}}$ and $S_{\mathrm{AB}}$ are the power spectral densities (PSD) of interferometer A and B and their CSD, respectively while $F_{\mathrm{A}}$, $F_{\mathrm{B}}$ and $F_{\mathrm{A}}^{\dagger}$, $F_{\mathrm{B}}^{\dagger}$ are the one-sided Fourier transforms and their complex conjugates and $N$ is the total number of spectra being averaged and $n$ is index of the spectra. The total number of spectra $N$, that result from an observation period $T$ is a function of the frequency bin-width ($f_{\textrm{BW}}$) of the Fast Fourier Transform (FFT) performed, which follow the relation $N=T\times f_{\textrm{BW}}$. For the data presented an $f_{\textrm{BW}}$ of 15.25\,kHz is chosen, which is the smallest bin-width that can be process by the real-time data acquisition system that has been commissioned.

The mean of the CSD follows a $\chi^{2}(N)$ distribution and for uncorrelated measurements like photon shot noise, the expectation value of the mean CSD scales as square root of the number of averages -

\begin{align}
    \langle S^{\mathrm{N}}_{\mathrm{AB}} \rangle &= \sqrt{\frac{S_{\mathrm{AA}}\cdot S_{\mathrm{BB}}}{N}} \label{eq:trend}
\end{align}

In the presence of a correlated signal between the two measurements, the expected mean tends towards the mean of the correlated signal while the variance scales down as the square root of the number of averages.

The output signal from the anti-symmetric ports of the interferometers is collected on to two Newfocus 1811-FS fast photo-detectors that have been modified to be able to measure up to 50\,mW of laser power, by increasing the active area to 1\,mm diameter and AC-coupling the front-end stage of
the circuit \cite{Grote_PD}. In steady-state operation, the fast photo-detectors are operated with about 30\,mW of laser power. Their frequency bandwidth is 80\,MHz and is currently limiting the measurement frequency band for the QUEST experiment. As part of an upgrade, a new set of fast photo-detectors with a bandwidth of 200\,MHz are being commissioned. Currently, the detection bandwidth of the instrument roughly covers the light-crossing frequency of the interferometer arms (81.97\,MHz)  and the main quantum gravity models being probed \cite{Kwon:2014yea,VERLINDE2021} predict a broadband signal peaked at this frequency. The planned 200\,MHz bandwidth of the instrument will help further constrain the nature of the signal being probed.

\begin{figure}[t]
    \includegraphics[width=8.6cm]{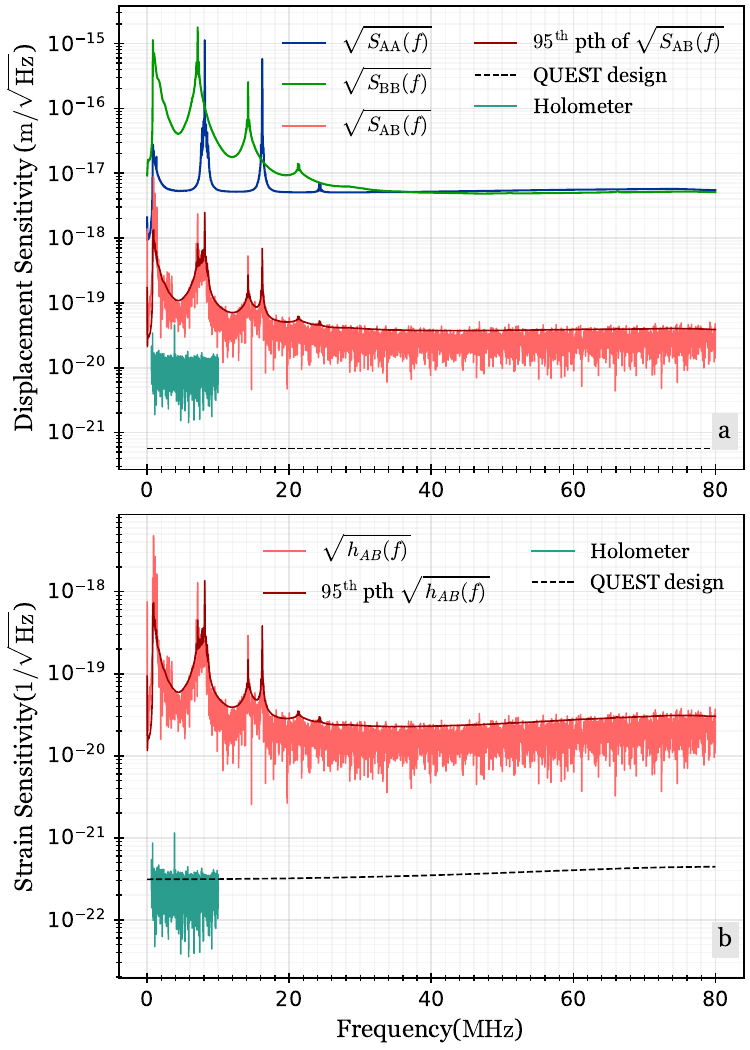}
    \caption{\label{fig:mainplot}a. The absolute displacement sensitivity of Interferometer-A $\sqrt{S_{\mathrm{AA}}(f)}$, Interferometer-B $\sqrt{S_{\mathrm{BB}}(f)}$  are shown in blue and green solid lines. The difference in their sensitivities at low frequencies is likely due to different laser frequency noise coupling to the anti-symmetric port of the interferometers. Special emphasis will be given to commission the interferometers and minimize the currently different macroscopic arm-asymmetries. Their CSDs $\sqrt{S_{\mathrm{AB}}(f)}$ averaged over the observation period is shown in light red and $\mathrm{95^{\mathrm{th}}}$ percentile of the averaged CSD is shown in solid red line. The CSD from the Holometer experiment \cite{Chou2017} is shown in cyan and the design sensitivity for QUEST experiment \cite{vermeulen_experiment_2021} is shown with dashed black line.\\
    b. The strain sensitivity derived from the the absolute displacement sensitivities and sky-position and polarization averaging are shown. The design strain sensitivity of the QUEST interferometers, being similar to that of the Holometer with table-top scale arms lengths, will require higher absolute displacement sensitivities. This will yield a better SNR accretion with time-averaging against the quantum gravity models being tested \cite{Kwon:2014yea,VERLINDE2021}. }
    
\end{figure}

The low-frequency technical noise content from the photo-detector signals is suppressed using analog high-pass filters with a corner frequency of 1\,MHz. The signals are then digitized with a resolution of 16 bits and at a rate of 500 million samples per second. The sampling rate and resolution product sets a data rate from digitizing two channels to 16 Gigabits per second. Storing timeseries data at this data rate is non-trivial. Averaging of frequency domain spectra provides a convenient method to compress the data to be stored, but at a cost of losing the ability to search for transient signals that are shorter in duration than the averaging time.

An FPGA platform from National Instruments consisting of two NI PXIe-5763 digitisers and a NI PXIe-7915 co-processor are used to implement a real-time data acquisition and processing system. For both channels that are sampled a window function is applied and an FFT is computed for time segments of a length of 32,786 samples, corresponding to 65.54\,$\mu$s of observed data. This sets the frequency resolution of the spectra to 15.250\,kHz. Without any loss of information, one-sided, complex-valued FFTs are propagated further to calculate two Power Spectral Densities (PSD) and one CSD. The mean of $10^{4}$ spectra are then calculated to compress the data from a total 0.655\,s of observation.  Thus averaged spectra are then written to a storage server as a `frame'.

Offline post-processing of frames, including vetoing of frames corresponding to lock-loss periods of the interferometers and correcting for the photo-detector response is done. The modified photo-detectors show a slight gain increase at 80\,MHz in their response, that is corrected for. Further averaging of frames is done to achieve the final displacement spectral density for an observation period.

The variance of the final averaged CSD is calculated using the relation

\begin{align}
    (\sigma_{\mathrm{AB}}^{2})_{N} = \frac{S_{\mathrm{AA}}(f)S_{\mathrm{BB}}(f)}{N}
\end{align}

\textit{Calibration and sensitivity -} The laser power transmitted through the end mirrors of the interferometers is used as an estimator for the power circulating in the arms. Then, the shot noise-dominated output of the interferometers is calibrated to displacement sensitivities using Eq.\,\ref{eq:sens}. The estimated calibration is propagated to the averaged PSDs obtained from the real-time data acquisition system, resulting in the spectra to be calibrated to displacement. A fiducial error of 15\% is assumed for the calibration as a result of technical uncertainties such as optical losses in the detection optics chain, quantum efficiencies of fast photo-detectors and uncertainties in the reflectivities of the core optics.

FIG.\,\ref{fig:mainplot}a shows the absolute displacement sensitivity in amplitude spectral density units of $\mathrm{m/\sqrt{Hz}}$, corresponding to interferometers A, B and the CSD between the interferometers. The spectra were obtained by averaging over a 10,000\,s of coincident observing run data. A short observing run was a strategic choice to evaluate the performance of the instrument, before proceeding with further commissioning of the interferometers to achieve even better sensitivity for longer observation length. 

Interferometers A and B are operated with an input power of 300\,mW each, from two separate Coherent Mephisto NPRO continuous-wave laser sources, that resulted in a circulating power on the beamsplitter of $83.32\pm1.02$\,W and $96\pm0.89$\,W, respectively. This resulted in absolute displacement sensitivities of about $5.5\times10^{-18}\,\mathrm{m/\sqrt{Hz}}$ in both interferometers. As part of this coincident observing run, a total of $1.64\times 10^{8}$ cross-spectra were collected and averaged, resulting in a cross-correlated displacement sensitivity of $3\times 10^{-20}\,\mathrm{m/\sqrt{Hz}}$.

The observed displacement sensitivity is transformed into strain sensitivity for stochastic gravitational waves that are assumed to be isotropic and unpolarized. The sky-position and polarization-averaged strain sensitivity $h(f)$ is calculated similar to LIGO \cite{schnabel2024opticalsensitivitiescurrentgravitational, Rakhmanov_2008}, and is shown in FIG.\,\ref{fig:mainplot}b.

\begin{figure}[t]
    \includegraphics[width=9cm]{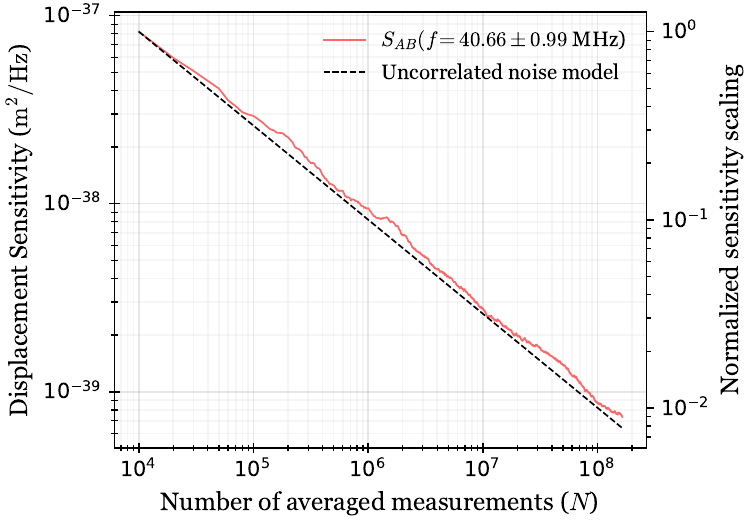}
    \caption{\label{fig:avgtrend} The averaging-scaling trend of the cross-correlated spectra $S_{\textrm{AB}}$ for the frequency bin 40.66$\pm$0.99\,MHz is shown in PSD units ($\textrm{m}^{2}/\textrm{Hz}$) and normalized sensitivity scaling. The expected trend for uncorrelated noise is also shown.}
\end{figure}

This broadband strain noise spectrum sets new upper limits for stationary length fluctuations in the frequency band 13 to 80\,MHz. We verified that the mean for the averaged cross-spectra follows a $\sqrt{N}$ trend for a growing number of averaged spectra $N$. Therefore, the signals between the interferometers are interpreted to be uncorrelated, shown in FIG. \ref{fig:avgtrend}.

This observation constrains stationary stochastic signal sources, such from the stochastic gravitational-wave background to the observed sensitivity.

Other interferometric experiments that have set limits at radio frequencies are the Holometer experiment \cite{Chou2017}, which set limits for the frequency band 1 to 13\,MHz with a strain sensitivity of about $4\times10^{-22}$  $\mathrm{1/\sqrt{Hz}}$ and a pair of co-located synchronous recycling interferometers \cite{Akutsu2008}, which set a narrow-band limit at 100\,MHz with a strain sensitivity of about $10^{-16}$  $\mathrm{1/\sqrt{Hz}}$. QUEST explores the frequency band between the two experiments and sets new limits from 13\,MHz to 80\,MHz with a strain sensitivity of $3\times10^{-20}$  $\mathrm{1/\sqrt{Hz}}$. Further commissioning is underway to improve both frequency band of observation and the strain sensitivity of the experiment.

\textit{Classical noise limit for a single interferometer -} Cross-correlation analysis and averaging can be performed with the output of a single interferometer to surpass the photon shot noise limited sensitivity \cite{PhysRevA.95.043831}. The expectation mean of the averaged CSDs will tend towards the technical, classical noises that limit the interferometer's sensitivity. 

\begin{figure}[t]
    \includegraphics[width=8.6cm]{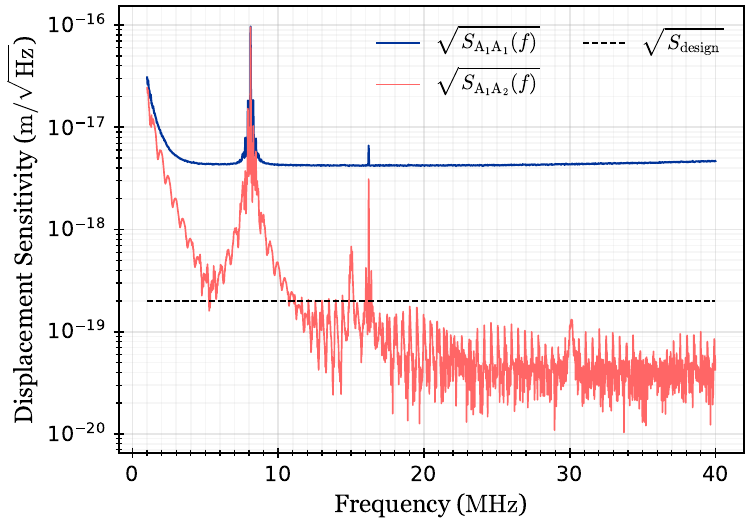}
    \caption{\label{fig:2pd} The two-photo-detector cross-correlation measurement, $\sqrt{\mathrm{S_{A_{1}A_{2}}}(f)}$, from the output of Interferometer-A is shown in red. The current observed shot noise limited sensitivity, $\sqrt{\mathrm{S_{A_{1}A_{1}}}(f)}$, is shown in blue and the shot noise-limited design sensitivity $\mathrm{\sqrt{S_{\mathrm{design}}}}$ of a single interferometer is shown in dashed black trace.}
\end{figure}

To demonstrate the functionality of our high-frequency cross-correlation pipeline and to analyze the noise sources below shot noise for a single interferometer, a single interferometer data collection run was conducted. A homodyne readout with a beamsplitter and two photo-detectors was set up at the output of interferometer-A. Each photo-detector sampled 50\,\% of the beam power leaving the interferometer, and a cross-correlation analysis between these photo-detectors was performed. The photon shot noise and photo-detector dark noise of each individual photo-detector are suppressed by the averaging of CSDs. About $1.4\times 10^{8}$ individual CSD spectra were collected and the resulting averaged spectrum is shown in FIG.\,\ref{fig:2pd}.

The classical noise shown at sub-shot noise sensitivities comprises laser frequency and laser intensity noises, thermo-refractive noise and bulk-thermal noise of the core optics. A characteristic comb of 470\,kHz lines and harmonics are seen as result of the thermal noise from the solid normal modes that arise from the 6\,mm thick, fused silica core optics used in the interferometer. This constitutes a unique thermal noise measurement of a fused silica optical element in this frequency range. The peaks at 8 and 16\,MHz are phase-modulation sidebands and their harmonics injected for length sensing and control of the interferometer.

\textit{Ongoing commissioning work -} The interferometers in QUEST are currently being further commissioned with the goal to achieve the design displacement sensitivity of $2\times10^{-19}$ $\mathrm{m/\sqrt{Hz}}$  \cite{vermeulen_experiment_2021}. The recycling factor of the power recycling cavity will be increased by installing new core optics with less optical losses (both scatter and absorption) and by increasing the reflectivity of the power recycling mirror to 99.9\%. The input power will be appropriately increased to reach the target circulating power of 10\,kW on the beamsplitters. For stable operation of the interferometers at high circulating powers, active feedback loops for the angular degrees of freedom are being commissioned. Automated re-locking protocols will be implemented into the digital control systems to maintain a high duty cycle of operation during observing runs.

New photo-detectors with a response bandwidth of 200\,MHz have been commissioned and are currently being characterized. These new photo-detectors will increase the detection bandwidth of the QUEST interferometers, and will allow the experiment to set new upper limits on strain sensitivities up to 200\,MHz.

To further improve the sensitivity of the interferometers, squeezed light injection along with a balanced homodyne readout scheme and Output Mode-Cleaners are currently being commissioned \cite{WillG2023}. An external calibration scheme to corroborate the shot noise based calibration and improve calibration uncertainties is being tested.

\textit{Conclusions -} The interferometers in the QUEST experiment have been commissioned to achieve absolute displacement sensitivities of $5.5\times10^{-18}$ $\mathrm{m/\sqrt{Hz}}$ and a coincident observing run was conducted to collect 10,000\,s of data. The observing run explores new territory by setting upper limits for stochastic gravitational waves in the frequency band 13 to 80\,MHz, at a strain sensitivity of $3\times10^{-20}$ $\mathrm{1/\sqrt{Hz}}$.

Further commissioning work for the experiment is ongoing, with the goal to achieve the design sensitivity and bandwidth of the experiment. An increase in the circulating power along with injection of squeezed states of light will target an absolute displacement sensitivity of each interferometer of $2\times10^{-19}$ $\mathrm{m/\sqrt{Hz}}$, and performing a 1 million second coincident observing run will allow for probing of correlated length fluctuation signals down to $4\times 10^{-22}$ $\mathrm{1/\sqrt{Hz}}$ strain sensitivity. This will result in about 2 orders of magnitude improvement on the constraint set in this paper along with covering a broader frequency band.

Operating the interferometers at their design sensitivities will allow us to probe for a wide variety of stationary signals like, signatures of quantised space-time, scalar field dark matter candidates and stochastic and continuous gravitational waves. Achieving high circulating powers in the interferometers will also aid in understanding the technical needs, especially in the context of limiting thermal noises, for the development and operation next generation of long-baseline, high-power gravitational wave detectors.

The authors are grateful for the support from the Science and Technology Facilities Council (STFC) Grants No. ST/T006331/1. and No. ST/W006456/1 for the Quantum Technologies for the Fundamental Physics program, and the Leverhulme Trust, Grant No. RPG-2019-022. HV is funded by Deutsche Forschungsgemeinschaft (DFG, German Research Foundation) under Germany’s Excellence Strategy – EXC-2123 QuantumFrontiers – 390837967. AP would like to thank STFC studentship Grant No. ST/W507374/1 and Cardiff University for providing support for the research work done.

\bibliography{apssamp}

\end{document}